\documentstyle[twoside,fleqn,npb,epsfig]{article}

\newcommand{\nn}{\nonumber}
\newcommand{\be}{\begin{equation}}
\newcommand{\ee}{\end{equation}}
\newcommand{\ba}{\begin{eqnarray}}
\newcommand{\ea}{\end{eqnarray}}

\def\gev{~{\rm GeV}}

\newcommand{\ov}[1]{\overline#1} 

\newcommand{\req}[1]{(\ref{#1})}

\def\xb{\bar{x}}
\def\sh{\hat{s}}
\def\uh{\hat{u}}
\def\={\,=\,}
\newcommand{\ci}[1]{\cite{#1}}

\newcommand{\AmS}{{\protect\the\textfont2
  A\kern-.1667em\lower.5ex\hbox{M}\kern-.125emS}}

\begin{document}
\title{Generalized Parton Distributions and wide-angle exclusive scattering} 

\author{P.\ Kroll \address{Fachbereich Physik, Universit\"at Wuppertal,\\ 
D-42097 Wuppertal, Germany\\
Email: kroll@physik.uni-wuppertal.de}}

\begin{abstract}
The handbag mechanism for wide-angle exclusive scattering reactions is 
discussed and compared with other theoretical approaches. Its application 
to  Compton scattering, meson photoproduction and two-photon annihilations 
into pairs of hadrons is reviewed.\\
Talk presented at 7th Zeuthen Workshop on Loops and Legs in Quantum
Field Theory
\end{abstract}

\maketitle

Recently a new approach to wide-angle Compton scattering off protons
has been proposed~\cite{rad98,DFJK1} where, for Mandelstam variables 
$s,-t,-u$ that are large as compared to a typical hadronic scale, $\Lambda^2$ 
of  the order of $1\gev^2$, the process amplitudes factorize into a hard 
parton-level subprocess, Compton scattering off quarks, and in soft form 
factors which represent $1/x$ moments of generalized parton distributions 
(GPDs) and encode the soft physics. Subsequently it has been realized
that this so-called handbag mechanism applies to a number of other
wide-angle reactions such as two-photon annihilations into pairs of hadrons
or meson photo- and electroproduction.

\begin{figure}[t]
\begin{center}
\includegraphics[width=3.2cm,bbllx=120pt,bblly=570pt,bburx=265pt,
bbury=680pt,clip=true]{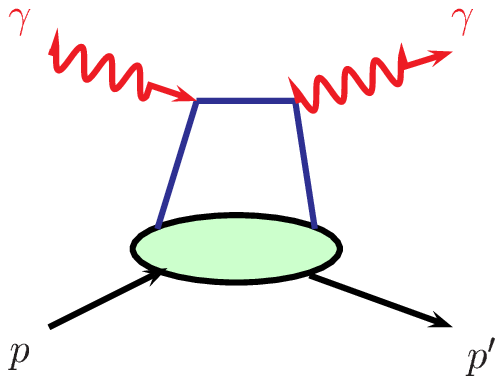} 
\includegraphics[width=3.2cm,bbllx=340pt,bblly=580pt,bburx=485pt,
bbury=693pt,clip=true]{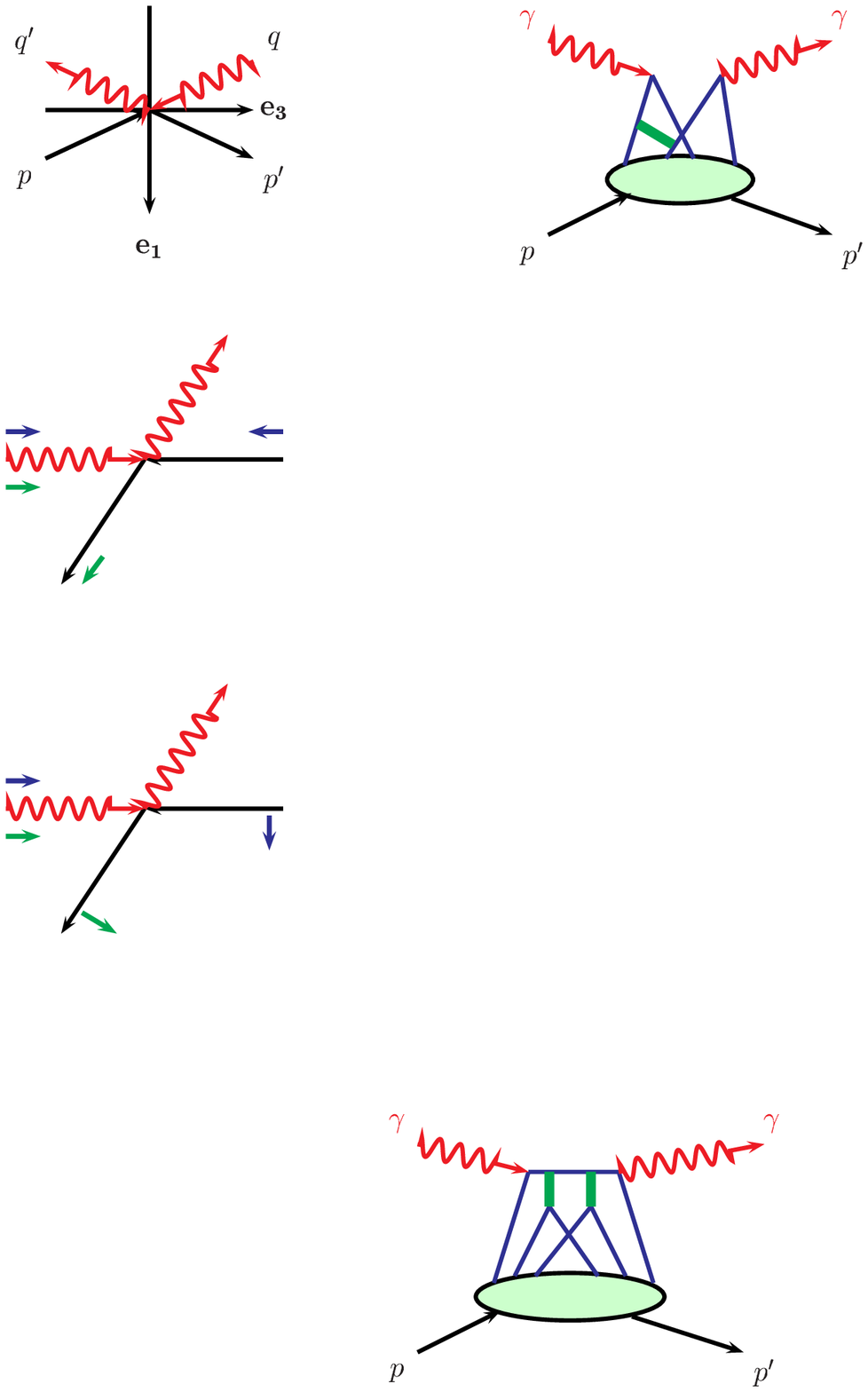} 
\vspace*{0.3cm} 
\includegraphics[width=3.0cm,bbllx=285pt,bblly=125pt,bburx=450pt,
bbury=240pt,clip=true]{hand-graph.ps} 
\includegraphics[width=3.6cm,bbllx=105pt,bblly=495pt,bburx=385pt,
bbury=610pt,clip=true]{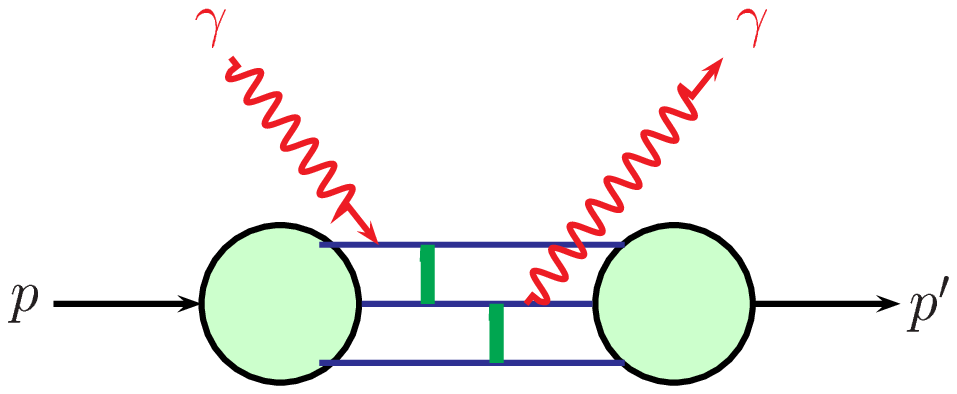} 
\caption{{}Handbag diagram for Compton scattering (upper left), cat's 
ears (upper right), the three-particle contribution (lower left) and
its valence quark approximation (lower left).}
\label{fig:handbag}
\end{center}
\vspace*{-1.2cm}
\end{figure}

There are competing mechanisms which contribute to wide-angle scattering 
besides the handbag which is characterized by one active parton, i.e.\
one parton from each hadron participates in the hard subprocess
(e.g.\ $\gamma q\to \gamma q$ in Compton scattering) while all others are
spectators. First there are the so-called cat's ears graphs (see Fig.\ 
\ref{fig:handbag}) with two active partons. 
It can be shown that in these graphs either a large parton virtuality
or a large parton transverse momentum occurs which forces the exchange
of at least one hard gluon. Hence, the cat's ears contribution is
expected to be suppressed as compared to the handbag one. The next
class of graphs are characterized by three active quarks and,
obviously, require the exchange of at least two hard gluons. In the
valence quark approximation which is expected to hold at large $-t$,
the big blob in the case of three active quarks decays into two
smaller blobs, see Fig.\ \ref{fig:handbag}. Each of these blobs
describe a hadron's distribution amplitude for finding valence quarks
in the hadron, each carrying some fraction $x_i$ of the hadron's
momentum.  This so-called leading-twist contribution    
is expected to dominate for asymptotically large momentum transfer but
is strongly suppressed for momentum transfer of the order of $10\, \gev^2$.
Since hadrons are not just made off their valence quarks one go on and 
consider four or more active partons. 
In principle, all the different contributions have 
to be added coherently. In practice, however, this is a difficult, 
currently impossible task since each contribution has its own
associated soft hadronic matrix element which, as yet, cannot be 
calculated from QCD. 
 
The contribution from the handbag diagram shown in
Fig.\ \ref{fig:handbag}, is calculated in a symmetrical frame which is 
a c.m.s.\ rotated in such a way that the momenta of the incoming ($p$) 
and outgoing ($p'$) proton momenta have the same light-cone plus 
components. Hence, the skewness defined as 
\be 
\xi \= \frac{(p - p')^{\,+}}{(p + p')^{\,+}}\,,
\ee
is zero. 
The crucial assumption in the handbag approach is that of restricted
parton virtualities, $k_i^2<\Lambda^2$, and of intrinsic transverse
parton momenta, ${\bf k_{\perp i}}$, defined with respect to their
parent hadron's momentum, which satisfy $k_{\perp i}^2/x_i
<\Lambda^2$, where $x_i$ is the momentum  fraction parton $i$ carries.   
 
One can then show \cite{DFJK1} that the subprocess Mandelstam variables
$\hat{s}$ and $\hat{u}$ are the same as the ones for the full process 
($s$ and $u$) up to corrections of order $\Lambda^2/t$. The active 
partons are approximately on-shell, move collinear with their parent 
hadrons and carry a momentum fraction close to unity, $x_j, x_j' \simeq 1$.
Thus, the physical situation is that of a hard parton-level subprocess
and a soft emission and reabsorption of quarks from the proton. The
arguments for handbag factorization hold in the time-like region as well 
\ci{DKV2} (see also Ref.\ \ci{weiss}). Here a suitable symmetrical frame 
is a c.m.s.\ in which the final state hadrons move in opposite directions 
along the 1-axis. Hence, $p^+=p'{}^+$ and the time-like skewness, defined as 
\be
\zeta\= \frac{p^+}{(p+p')^{\,+}}\,,
\ee
is $1/2$. 
 
The light-cone helicity amplitudes for wide-angle Compton scattering
read \ci{DFJK1,HKM} 
\ba
\label{ampl}
\lefteqn{{M}_{\mu'\nu',\,\mu \nu}(s,t) = \; \frac{e^2}{2}\, 
          \Big[\, \delta_{\nu'\nu}\,{ T}_{\mu'\nu,\,\mu\nu}\, (R_V + R_A) } \nn\\[0.3em] 
       &+& \, \delta_{\nu'\nu}\,{ T}_{\mu'-\nu,\,\mu -\nu}\, 
                        (R_V - R_A)  \\[0.3em]
  &+& \,\delta_{-\nu'\nu}\,\frac{\sqrt{-t}}{2m} 
         \left(\,  T_{\mu'-\nu',\,\mu\nu}\,  
   + \, { T}_{\mu'\nu',\,\mu -\nu}\, \right) \,R_T \Big] \,. \nn 
\ea
The amplitudes for other wide-angle reaction have an analogue structure.
$\mu (\nu)$ and $\mu'(\nu')$ denote the helicities of the incoming and 
outgoing photons (protons in $M$ or quarks in the subprocess amplitude
$T(\sh,t)$), respectively. $m$ denotes the mass of the proton. The form 
factors $R_i(t)$  represent $1/\xb$-moments of GPDs at zero skewness. 
For Compton scattering the hard scattering has been calculated to 
next-to-leading order perturbative QCD \cite{HKM}. To this order one
has to take into account the photon-gluon subprocess and a corresponding 
gluonic form factor. This small correction which amounts to less than
$10\%$ in the cross section, is taken into account in the numerical 
results shown below but, for convenience, ignored in the formulas.  

The handbag amplitude \req{ampl} leads to the following leading-order 
Compton cross section  
\ba
\frac{d\sigma}{dt} &=& \frac{d\hat{\sigma}}{dt} \left\{ \frac12\, \big[
R_V^2(t)\,(1+\kappa_T^2) + R_A^2(t)\big] \right.\nn\\
&&\hspace*{-0.5cm}\left.  - \frac{\uh\sh}{\sh^2+\uh^2}\, \big[R_V^2(t)\,(1+\kappa_T^2) 
                - R_A^2(t)\big]\right\}\,,
\label{dsdt}
\ea
where $d\hat{\sigma}/dt$ is the Klein-Nishina cross section for
Compton scattering off massless, point-like spin-1/2 particles of
charge unity. The quantity $\kappa_T$ is defined as 
\be
\kappa_T\= \frac{\sqrt{-t}}{2m}\, \frac{R_T}{R_V}\,. 
\label{kappaT}
\ee
Another interesting observable in Compton scattering is the helicity
correlation, $A_{LL}$,  between the initial state photon and proton
or, equivalently, the helicity transfer, $K_{LL}$, from the incoming
photon to the outgoing proton. In the handbag approach one obtains
\cite{DFJK1,HKM} 
\be
A_{LL}\=K_{LL}\simeq \frac{\sh^2 - \uh^2}{\sh^2 + \uh^2}\, 
                    \frac{R_A(t)}{R_V(t)}\,,
\label{all}
\ee  
where the factor in front of the form factors is the corresponding
observable for $\gamma q\to \gamma q$. The result \req{all} is a
robust prediction of the handbag mechanism, the magnitude of the
subprocess helicity correlation is only diluted by the ratio of the 
form factors $R_A$ and $R_V$. 

The cross section for two-photon annihilations into
baryon-antibaryon pairs reads~\ci{DKV2} 
\ba
\lefteqn{\frac{d\sigma}{dt}\,(\,\gamma\gamma\,\to\,\, B\ov{B}\,) \= 
      \frac{4\pi\alpha^2_{\rm elm}} 
        {s^2 \sin^2 \theta} \Big\{ \big|R_A^B(s)+ R_P^B(s)\big|^2}\nn\\
     &+& \cos^2\theta\,\big|R_V^B(s)\big|^2\,+\, 
                     \frac{s}{4m^2}\,\big| R_P^B(s)\big|^2 
                                       \Big\}\,,
\label{ann-BB}
\ea
while for pseudocalar meson it is given by
\be
\frac{d\sigma}{dt}(\gamma\gamma\to M\ov{M}) \= 
     \frac{8\pi\alpha^2_{\rm elm}} 
              {s^2 \sin^4 \theta} \big|R_{M\ov{M}}(s)\big|^2\,.
\label{ann-MM}
\ee
The annihilation form factors represent moments of two-hadron
distribution amplitudes, $\Phi_{2h}(z,\zeta,s)$, which are time-like
versions of GPDs.  

In order to make actual predictions for Compton scattering a model for 
the form factors or rather for the underlying GPDs is required. A  first 
attempt to parameterize the GPDs $H$ and $\widetilde{H}$ at zero
skewness is~\cite{rad98,DFJK1} 
\ba
H^{\,q}(\xb,0;t) &=& \exp{\left[a^2 t
        \frac{1-\xb}{2\xb}\right]}\, q(\xb)\,,\nn\\ 
\widetilde{H}^{\,q}(\xb,0;t) &=& \exp{\left[a^2 t
        \frac{1-\xb}{2\xb}\right]}\, \Delta q(\xb)\,,
\label{gpd}
\ea
where $q(\xb)$ and $\Delta q(\xb)$ are the usual unpolarized and 
polarized parton distributions in the proton. The transverse size of
the proton, $a$, is the only free parameter. The model (\ref{gpd}) is
designed for large $-t$. Hence, forced by the exponential in (\ref{gpd}), 
large $\xb$ is implied, too. Despite this the normalizations of the 
model GPDs at $t=0$ are correct. 
 
\begin{figure}[t]
\begin{center} 
\includegraphics[width=5.2cm,bbllx=100pt,bblly=74pt,bburx=584pt,bbury=631pt,%
angle=-90,clip=true]{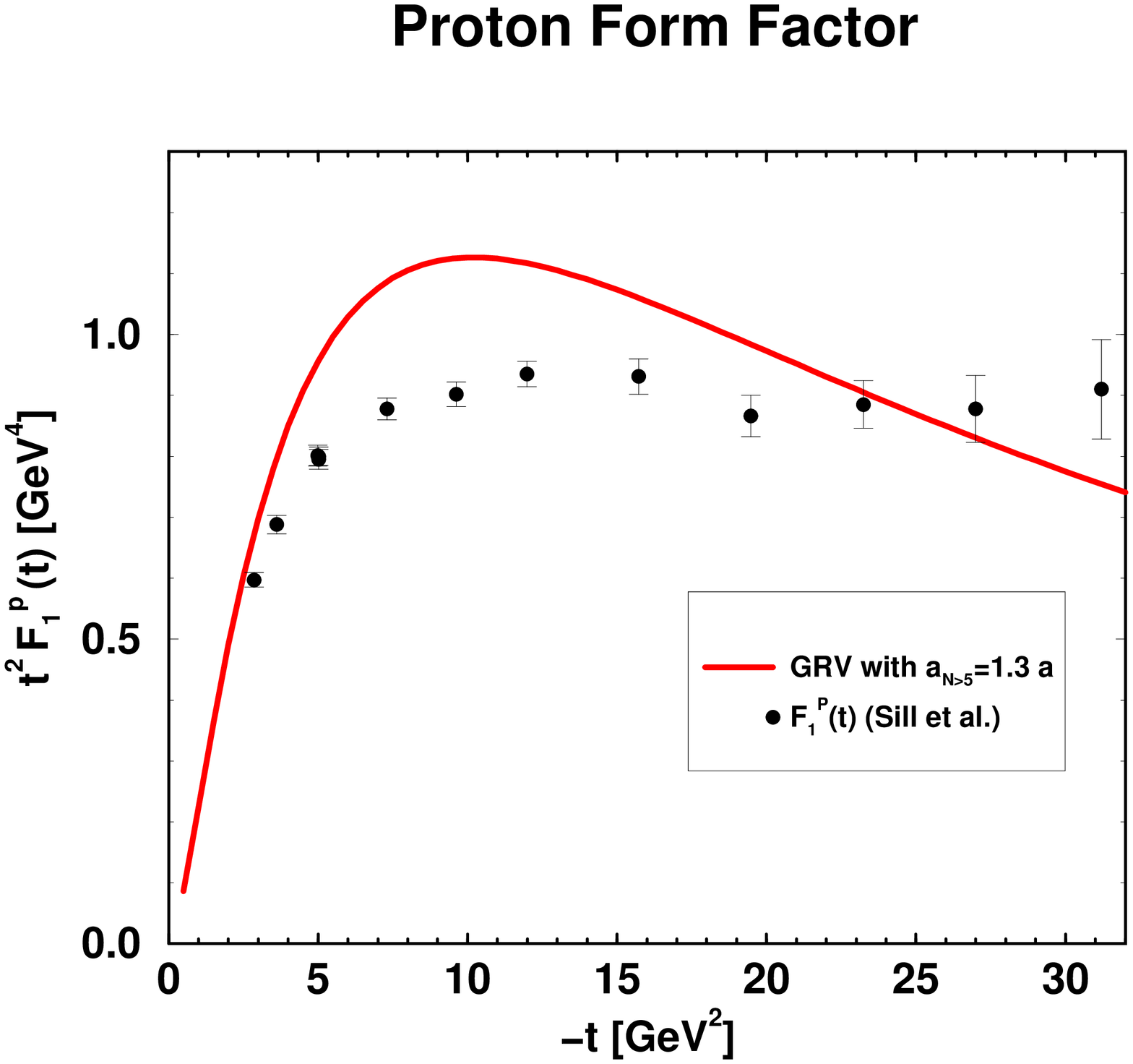} 
\includegraphics[width= 6.0cm,bbllx=81pt,bblly=49pt,bburx=398pt, 
bbury=293pt,clip=true]{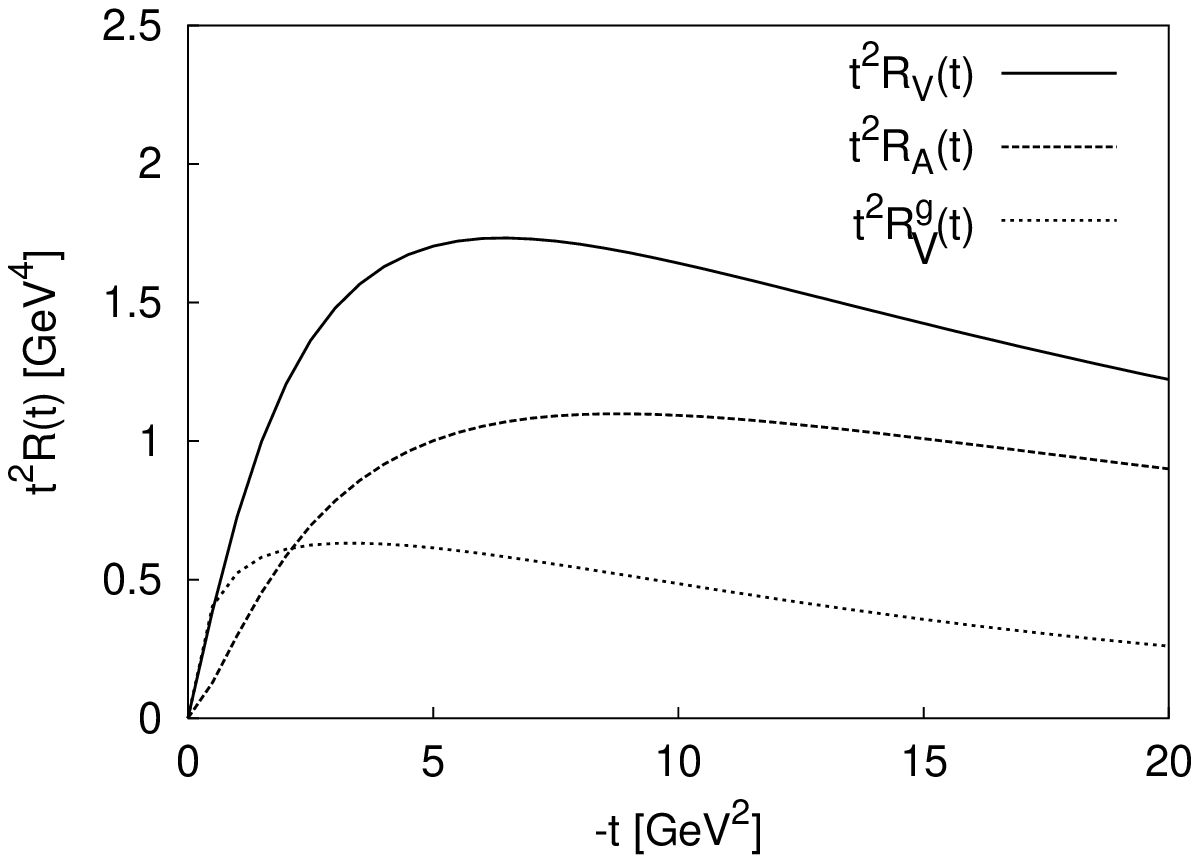} 
\vspace*{-0.5cm}
\caption{Model results for the scaled Dirac form factor of the proton 
  $t^2F_1^p$ (top) and the Compton form factors $t^2R_i$ (bottom) 
  (in $\gev^4$). Data taken from Ref.\
  \protect\cite{sill}.}   
\label{fig:form}
\end{center}
\vspace*{-1.4cm}
\end{figure}

From the model GPDs \req{gpd} we can evaluate the various form
factors by taking appropriate moments. For the Dirac and the axial form 
factor one has the sum rules
\ba
F_1(t) &=&\sum_q e_q \int_{-1}^1 d\xb H^{\,q}(\xb,0;t)\,, \nn\\
F_A(t) &=& \int_{-1}^1 d\xb \Big[ \widetilde{H}^{\,u}(\xb,0;t)-
             \widetilde{H}^{\,d}(\xb,0;t)\Big],
\label{formfactors}
\ea
while the Compton form factors read
\ba
R_V(t) &=&\sum_q e_q^2 \int_{-1}^1 \frac{d\xb}{\xb} H^{\,q}(\xb,0;t)\,, \nn\\
R_A(t) &=& \sum_q e_q^2 \int_{-1}^1 \frac{d\xb}{\xb} {\rm sign}(\xb)\, 
\widetilde{H}^{\,q}(\xb,0;t).
\label{Compton-formfactors}
\ea
Numerical results for the proton's Dirac form factor and the Compton
form factors are shown in Fig.\ \ref{fig:form}. The scaled form factors 
$t^2 F_{1,A}$ and $t^2 R_i$ exhibit broad maxima which mimick
dimensional counting in a range of $-t$ from, say, $5$ to about
$20\,\gev^2$. As the comparison with experiment~\ci{sill} reveals, 
the model GPDs work rather well although the predictions for the Dirac 
form factor overshoot the data by about $20 - 30\,\%$ for $-t$ around 
$5 \gev^2$. An effect of similar size can be expected for the Compton 
form factors. A phenomenological analysis of the GPDs is currently 
in progress. It bases on a more general ansatz for the general profile 
function in \req{gpd} with a few free parameters which are adjusted to 
the data on the nucleon form factors through the sum rules. In this
analysis the GPD $E^q$ being related to nucleon helicity flip, will 
also be determined. With it at our disposal one can evaluate the tensor 
form factor $R_T$ analogously to \req{Compton-formfactors}. In the
present stage only predictions for various constant values of the 
quantity $\kappa_T$ \req{kappaT} are given.

It is important to realize that the GPDs represent process independent
information on the nucleon. They therefore appear for instance in
meson photoproduction as well. However, the flavor composition of the
full form factors as well as the relative sign between quark and
antiquark contributions are process dependent. In time-like processes 
form factors occur which are analytic continuations of the above 
space-like ones. Thus, for instance, in $\gamma\gamma \to p\ov{p}$ we
have 
\be
R_i(s) \= \sum_q e_q^2 F_{i}^{\,q}(s) \,,    \qquad i=V,A,P
\ee
with
\be 
F_i^{\,q}(s) \= \int dz \Phi_{p\ov{p}\,i}^{\,q} (z,\zeta=1/2,s) \,.
\ee
In contrast to the space-like region where the pseudoscalar form
factor decouples in the symmetric frame, here the scalar form factor
does not contribute. The vector form factor is therefore related to
the magnetic one
\be
G_M(s) \= \sum_q e_q F_V^{\,q}(s)\,,
\ee
and not to the Dirac form factor. It is expected that the
time-like form factors are of the same order of magnitude than the
space-like ones.
   
Employing the model GPDs and the corresponding form factors, various
Compton observables can be calculated~\cite{DFJK1,HKM}. The
predictions for the differential cross section are in fair agreement
with the Cornell data~\cite{shupe}. 
The JLab E99-114 collaboration~\cite{nathan} 
will provide accurate cross section data soon which will allow for a 
crucial examination of the handbag mechanism.

\begin{figure}[t]
\begin{center}
\includegraphics[width=6.5cm,bbllx=78pt,bblly=52pt,bburx=400pt,bbury=294pt, 
clip=true]{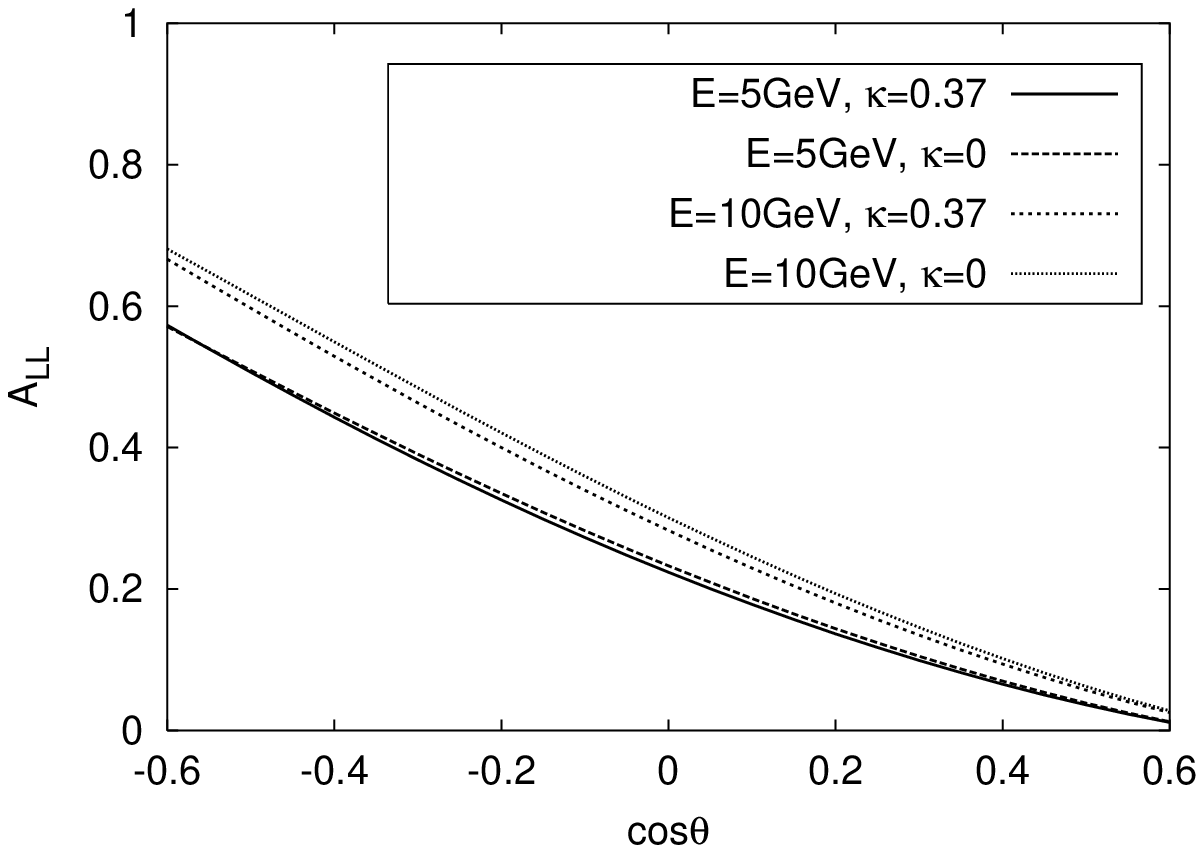} 
\includegraphics[width=6.5cm,bbllx=159pt,bblly=448pt,bburx=475pt, 
bbury=664pt,clip=true]{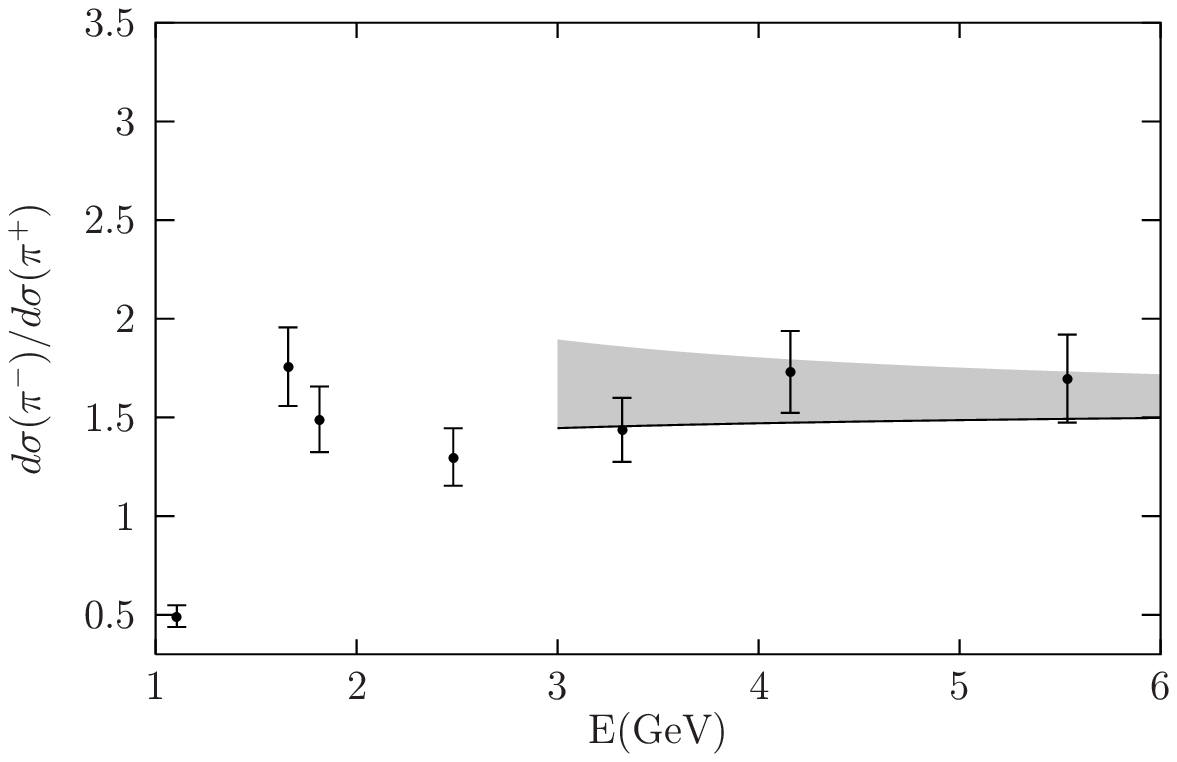}
\vspace{-1.0cm} 
\caption{Predictions for the helicity correlation $A_{LL}=K_{LL}$ in
Compton scattering (top) and the ratio of the $\gamma n\to \pi^- p$ and
$\gamma p\to\pi^+ n$ cross sections vs.\ beam energy, $E$, at a
c.m.s. scattering angle of $90^\circ$ (bottom). Data taken from \ci{zhu}.}
\label{fig:cross}
\end{center}
\vspace*{-0.9cm}
\end{figure}

Predictions for $A_{LL}=K_{LL}$ are shown in Fig.\ \ref{fig:cross}.
The JLab E99-114 collaboration \cite{nathan} has presented a first
measurement of $K_{LL}$ at a c.m.s.\ scattering angle of
$120^\circ$ and a rather low photon energy of $3.23 \gev$. This still
preliminary data point is in fair agreement with the predictions
from the handbag given the small energy at which they are
available. 

For photo- and electroproduction of mesons the dynamics of the
subprocess, $\gamma q \to M q$, is to be specified. The simplest
mechanism is the one-gluon exchange. It turns out that this
contribution fails with the normalization of the photoproduction cross
section \ci{hanwen}. Treating the subprocess in a more general way by
a covariant decomposition and neglecting quark helicity flip, one 
obtains a number of predicitions whose experimental verification would
signal the dominance of the handbag mechanism. Thus, for instance, 
the helicity correlation $A_{LL}$ is similar to the result \req{all}
for Compton scattering. Another interesting result is the ratio of the
cross sections for photoproduction of $\pi^+$ and $\pi^-$. It
is approximately given by 
\be
\frac{d\sigma(\gamma n\to \pi^- p)}{d\sigma(\gamma p\to \pi^+ n)} \simeq
\left[\frac{e_d \uh + e_u \sh}{e_u \uh + e_d \sh}\right]^2\,.
\label{pi-ratio}
\ee
The form factors cancel in the ratio. The prediction~\req{pi-ratio} is
in fair agreement with a recent JLab measurement \cite{zhu}, see 
Fig.\ \ref{fig:cross}. 

\begin{figure}[t]
\begin{center}
\includegraphics[width=5.5cm,bbllx=73pt,bblly=22pt,bburx=527pt,bbury=429pt,clip=true]
{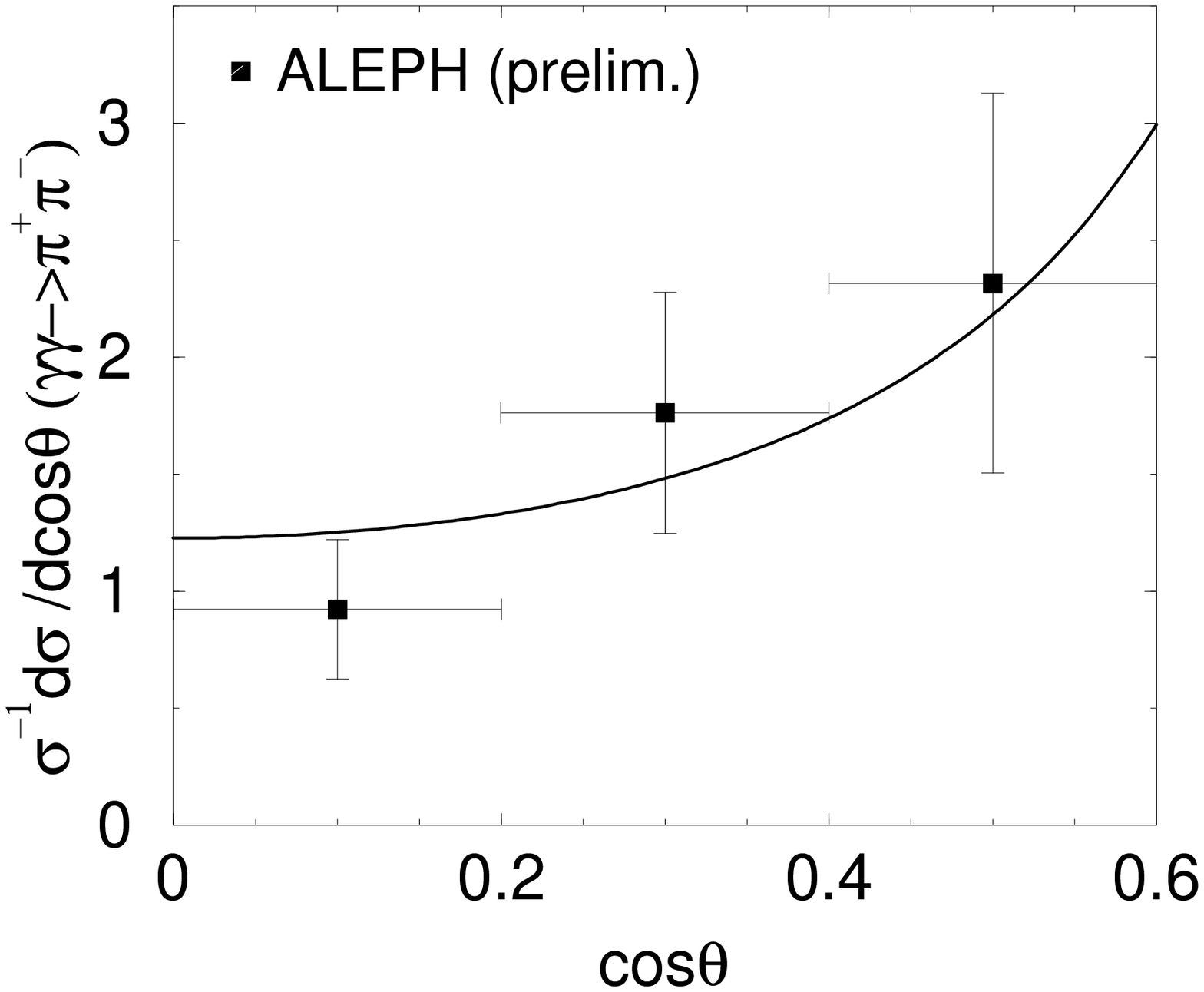}
\vspace*{-0.9cm}
\caption{Handbag predictions for the angular dependence of the cross section
for $\gamma\gamma\to \pi^+\pi^-$ versus $s$.  
Preliminary data are taken from ALEPH~\protect\cite{ALEPH}.} 
\label{fig:pipi}
\end{center}
\vspace*{-0.9cm}
\end{figure}

The characteristic $\sin^{-4}{\theta}$ dependence of the cross section
for two-photon annihilation into a pair of pseudoscalar mesons
\req{ann-MM} is predicted in fair agreement with experiment, see 
Fig.\ \ref{fig:pipi}. The form factors have not been modelled in 
Refs.\ \cite{DKV2} but rather extracted ('measured') from the 
experimental cross section. The average value of the scaled form factor 
$sR_{2\pi}$ obtained that way is $0.75\gev^2$. The closeness of this
value to that of the scaled time-like electromagnetic form factor of the 
pion ($0.93 \pm 0.12 \gev^2$) hints at the internal consistency of the 
handbag approach. Similar results are found for the production of 
baryon-antibaryon pairs. 

Another feature of the handbag mechanism in the time-like
region is the intermediate $q\ov{q}$ state implying the absence of 
isospin-two components in the final state. A consequence of this property is 
\be
\frac{d\sigma}{dt}(\gamma\gamma\to \pi^0\pi^0) = 
                 \frac{d\sigma}{dt}(\gamma\gamma\to \pi^+\pi^-)\,,
\ee
which is independent of the soft physics input and is, in so far, 
a robust prediction of the handbag approach. The absence of the
isospin-two components combined with flavor symmetry allows one to
calculate the cross sections for other $B\ov{B}$ channels using the 
$p\ov{p}$ form factors as the only soft physics input. 

To summarize, I have briefly reviewed the theoretical activities on
applications of the handbag mechanism to wide-angle scattering. There
are many interesting predictions, some are in fair agreement with
experiment, others still awaiting their experimental examination. It
seems that the handbag mechanism plays an important role in exclusive
scattering for momentum transfers of the order of $10\gev^2$. However,
before we can draw firm conclusions more experimental tests are needed.
I finally emphasize that the structure of the handbag amplitude,
namely its representation as a product of perturbatively calculable
hard scattering amplitudes and $t\, (s)$-dependent form factors is the
essential result. Refuting the handbag approach necessitates  
experimental evidence against this factorization.

Acknowledgments: It is a pleasure to thank J.\ Bl\"umlein, S.\ Moch and T.\
Riemann for organising the interesting meeting in Zinnowitz.


\begin{thebibliography}{99}
\bibitem{rad98} A.~V.~Radyushkin,
Phys.\ Rev.\ D {\bf 58}, 114008 (1998).

\bibitem{DFJK1} M.~Diehl, T.~Feldmann, R.~Jakob and P.~Kroll,
Eur.\ Phys.\ J.\ C {\bf 8}, 409 (1999),
Phys.\ Lett.\ B {\bf 460}, 204 (1999).

\bibitem{DKV2} M.~Diehl, P.~Kroll and C.~Vogt,
Phys.~Lett.~{\bf B} 532, 99 (2002),  
Eur.\ Phys.\ J.\ C {\bf 26}, 567 (2003).

\bibitem{weiss} A.~Freund, A.~V.~Radyushkin, A.~Schafer and C.~Weiss,
Phys.\ Rev.\ Lett.\  {\bf 90}, 092001 (2003). 

\bibitem{HKM} H.~W.~Huang, P.~Kroll and T.~Morii,
Eur.\ Phys.\ J.\ C {\bf 23}, 301 (2002).

\bibitem{sill} A.F.\ Sill {\it et al.}, Phys.\ Rev.\ D{\bf 48}, 29
(1993); A.\ Lung {\it et al.}, Phys.\ Rev.\ Lett.\ {\bf 70}, 718 (1993).


\bibitem{shupe} M.A.\ Shupe {\em et~al.},
                Phys. Rev. {\bf D19}, 1921 (1979).

\bibitem{nathan} E99-114 JLab collaboration,
spokespersons C.\ Hyde-Wright, A.\ Nathan and B.\ Wojtsekhowski.


\bibitem{hanwen} H.~W.~Huang and P.~Kroll,
Eur.\ Phys.\ J.\ C {\bf 17}, 423 (2000);
H.~W.~Huang, R.~Jakob, P.~Kroll and K.~Passek-Kumericki,
Eur.\ Phys.\ J.\ C {\bf 33}, 91 (2004).


\bibitem{zhu} L.~Y.~Zhu {\it et al.}  [Jefferson Lab Hall A Collaboration],
Phys.\ Rev.\ Lett.\ {\bf 91}, 022003 (2003).

\bibitem{ALEPH} A.\ Finch [for the ALEPH collaboration], Proceedings of
the {\it International Conference on The Structure and Interactions of 
the Photon (PHOTO N2001)}, Ascona 2001 (World Scientific, Singapore).


\end{thebibliography}
\end{document}